\documentclass[aps,prl,twocolumn,superscriptaddress,floatfix]{revtex4}
\usepackage{epsfig,amssymb}
\usepackage{color}
\begin{document}

\author{F. Heidrich-Meisner}
\author{A. Honecker}\author{D.C. Cabra}\author{W. Brenig}
\affiliation{Technische Universit\"at Braunschweig, Institut f\"ur
  Theoretische Physik, Mendelssohnstr.\ 3, 38106 Braunschweig,
  Germany}


\date{February, 12th, 2004}

{\bf Comment on ``Anomalous Thermal Conductivity of Frustrated
Heisenberg Spin Chains and Ladders''\\}

\begin{figure}[th]
\centerline{\epsfig{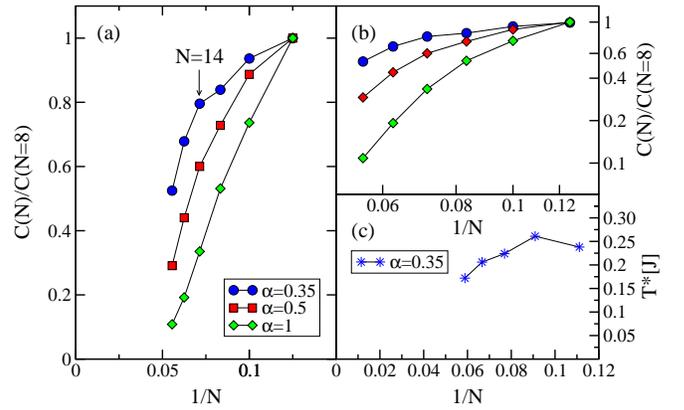}}
\caption{(a), (b) Size-dependence ($N=8,10,12,14,16,18$) of the
high-temperature residue $C(N)$ 
for different values of the frustration $\alpha$ in the gapped
regime. (b) Both axes are scaled logarithmically.  (c) Size
dependence of $T^*(N^*)$ (see text for details) for
$\alpha=0.35$. Solid lines are guides to the eye.\label{fig:1}}
\end{figure}

\indent In a recent Letter \cite{gros02}, Alvarez and Gros have
presented a numerical study of the thermal conductivity  of frustrated chains and
spin ladders with spin $1/2$. Using exact diagonalization of finite systems with
$N\leq 14$ sites, they have computed the zero-frequency weight
$\kappa^{\mathrm{(th)}}(T,N)$, i.e., the Drude weight, of the
thermal conductivity where $T$ is the temperature and $N$ the number of sites.
One of their main conclusions is that  the numerical data 
indicate  a {\it finite} value of $\kappa^{\mathrm{(th)}}$ in
the thermodynamic limit ($N\to \infty$) for 
spin ladders and frustrated chains.
In the latter case, this conclusion is based
on a finite-size analysis of the high-temperature residue $C(N)$,
 given by $\lim_{T\to\infty}\lbrack T^2\kappa^{\mathrm{(th)}}(T,N)\rbrack=C(N)$,
 for $\alpha=0.1,0.24,0.35$ (see
Refs.\ \cite{gros02,hm02} for definitions; also note \cite{factorofpi}).\\
\indent In this Comment, we argue that, from the systems investigated in
\cite{gros02}, {\em no} conclusions of a finite thermal Drude weight
in the gapped regime of frustrated chains for $N\to \infty$ can be drawn.
This will be corroborated by supplementary data for systems up to $N=18$ sites.
In Figs.~\ref{fig:1}(a) and \ref{fig:1}(b), we show the size
dependence of $C(N)$  for $8\leq N\leq 18$ and
$\alpha=0.35,0.5,1$ where $C(N)/C(N=8)$ is plotted versus $1/N$.
Figure \ref{fig:1}(b) is a log-log display of Fig.~\ref{fig:1}(a). An overall
and monotonic decrease of $C(N)$ is evident, consistent
with a {\em vanishing} thermal Drude weight for $T\gg J$ and
$N\to \infty$. Indeed, the curvature of the curves in
the log-log plot of $C(N)/C(N=8)$ versus $1/N$
suggests that $C(N)$ vanishes more rapidly than any power of $1/N$
as $N\to\infty$. Most important,  for the case of $\alpha=0.35$
and $8\leq N\leq 14$ studied in \cite{gros02}, the overall
decrease of $C(N)$ remains clearly observable, except for minor
finite-size oscillations, which, however, justify no extrapolation to a
finite value for $N\to \infty$.\\
\indent In their Letter, Alvarez and Gros have also argued that the
behavior of $\kappa^{\mathrm{(th)}}(T,N)$ for $\alpha=0.35$ at {\it
low} temperatures supports the conclusion of a finite thermal
Drude weight in the thermodynamic limit. Indeed, there is a crossover
temperature $T^*(N^*)$ \cite{Tstar} where the monotonic decrease of
$\kappa^{\mathrm{(th)}}(T,N)$ with increasing system size
observed at high temperatures changes to a monotonic increase of
$\kappa^{\mathrm{(th)}}(T,N)$ with system size (see Fig.\ 3 in Ref.\
\cite{hm02}). However, as shown in  Fig.~\ref{fig:1}(c), $T^*(N^*)$ decreases with increasing system size already for
$N^{*}\geq 11$, i.e., {\em including}  systems studied in \cite{gros02},
and could well extrapolate to zero for $N^*\to \infty$. In any case,
finite-size effects for $T\lesssim 0.1J$ are far too large even at
$N=18$ to allow for any reliable predictions regarding
$\kappa^{\mathrm{(th)}}(T)$ in the thermodynamic limit.\\
\indent 
 Summarizing our numerical analysis of gapped, frustrated chains,
both in the high- and the low-temperature regime we find
no evidence in favor of a finite thermal Drude weight in
the thermodynamic
limit in contrast to  Ref.~\cite{gros02}.\\
\indent Finally, for spin ladders Alvarez and Gros have used only the size
dependence of $\kappa^{\mathrm{(th)}}(T,N)$ at low temperatures
to conjecture a finite $\kappa^{\mathrm{(th)}}(N\to \infty)$.  In view
of the results for the frustrated chain presented in this Comment, we suggest
to perform a finite-size analysis of the high-temperature residue $C(N)$
to substantiate this conjecture. We believe that this is important
for the analysis of the thermal conductivity in the spin ladder
 material La$_5$Ca$_9$Cu$_{24}$O$_{41}$\cite{hess01}.\\
\indent We acknowledge financial support by the DFG 
and a DAAD-ANTORCHAS exchange program. \\

\noindent
F. Heidrich-Meisner${}^{1}$, A. Honecker${}^{1}$, D.C. Cabra${}^{2}$, W. Brenig${}^{1}$\\
\noindent
${}^1$Institut f\"ur Theoretische Physik, TU Braunschweig,\\
38106 Braunschweig, Germany\\
${}^2$Laboratoire de Physique, ENS-Lyon, 69364 Lyon, France\\ 



\begin{thebibliography}{99}
\bibitem{factorofpi} For $N\leq 14$ and
$\alpha=0.35$ the data in Ref.\ \cite{gros02} are identical to
that published in \cite{hm02} which, however, includes systems up
to $N= 18$. The definitions of $\kappa^{\mathrm{(th)}}$ in 
Refs.\ \cite{gros02,hm02} differ by a trivial factor of $\pi$.
\bibitem{Tstar} We define $T^*(N^*)$ by $\kappa^{\mathrm{(th)}}(T^*,N+2)
=\kappa^{\mathrm{(th)}}(T^*,N)$  with $N^*=(N+1)$ and even $N$.
\end{thebibliography}
\end{document}